# QED for fields obeying a square root operator equation

Tobias Gleim

Instead of using local field equations – like the Dirac equation for spin-1/2 and the Klein-Gordon equation for spin-0 particles – one could try to use non-local field equations in order to describe scattering processes. The latter equations can be obtained by means of the relativistic energy together with the correspondence principle, resulting in equations with a square root operator. By coupling them to an electromagnetic field and expanding the square root (and taking into account terms of quadratic order in the electromagnetic coupling constant $e$), it is possible to calculate scattering matrix elements within the framework of quantum electrodynamics, e.g. like those for Compton scattering or for the scattering of two identical particles. This will be done here for the scalar case. These results are then compared with the corresponding ones based on the Klein-Gordon equation. A proposal of how to transfer these reflections to the spin-1/2 case is also presented.

Free scalar particles are usually described by means of the well-known Klein-Gordon equation (see e.g. [4,5,6,10]):

$$\left(\partial_t^2 + \hat{\vec{p}}^2 + m^2\right)\phi(\vec{x},t) = 0, \tag{1}$$

where we have used the momentum operator in configuration space $\hat{\vec{p}} = -i\vec{\nabla}$ (and set the velocity of light as well as Planck's constant $\hbar$ to one). (1) can be regarded as an iteration of the following square root operator equation (see e.g. [1,2,4,5]):

$$i\partial_t \phi(\vec{x},t) = \left(\sqrt{m^2 + \hat{\vec{p}}^2}\,\phi\right)(\vec{x},t). \tag{2}$$

Introducing an electromagnetic field with a 4-vector potential $(A^\mu) = (A^0, \vec{A})$ and applying minimal coupling (with coupling constant $e$), i.e. replacing $i\partial_\mu = i\partial/\partial x^\mu$ by $i\partial/\partial x^\mu - eA^\mu(x)$, the Klein-Gordon equation yields (see e.g. [6])

$$\left(\partial_t^2 + \hat{\vec{p}}^2 + m^2\right)\phi(x) = \left[-ie\left(\partial_\mu A^\mu(x) + A^\mu(x)\partial_\mu\right) + e^2 A_\mu(x)A^\mu(x)\right]\phi(x), \tag{3}$$

where we have used the 4-vector notation $(x^\mu) = (x^0, \vec{x}) = (t, \vec{x})$ and Einstein's summation convention. Here, the coupling terms on the right hand side of (3) could easily be separated from the term with the free particle Hamiltonian on the left hand side of (3). This is unfortunately no longer possible, if one couples the non-local equation (2) to the electromagnetic field:

$$i\partial_t \phi(x) = \left(\sqrt{m^2 + \left(\hat{\vec{p}} - e\vec{A}(x)\right)^2}\,\phi\right)(x) + eA^0(x)\phi(x), \tag{4}$$

because the vector potential $\vec{A}$ appears under the square root. But in a perturbation analysis of scattering processes, this property is useful, since such an analysis is based on the assumption that the coupling terms make only small contributions to the free particle solution due to the small value of the coupling constant $e$. By rewriting the Hamiltonian in (4),

$$\hat{H}' = eA^0 + \sqrt{m^2 + \left(\hat{\vec{p}} - e\vec{A}(x)\right)^2} = eA^0 + \sqrt{\left(m^2 + \hat{\vec{p}}^2\right) + \left(-e\hat{\vec{p}}\cdot\vec{A} - e\vec{A}\cdot\hat{\vec{p}} + e^2\vec{A}^2\right)}, \tag{5}$$

one can split off a factor with the free Hamiltonian

$$\hat{H}'_0 = \sqrt{m^2 + \hat{p}^2}, \tag{6}$$

which yields

$$\hat{H}' = eA^0 + \sqrt{1 + \left(-e\hat{\vec{p}}\cdot\vec{A} - e\vec{A}\cdot\hat{\vec{p}} + e^2\vec{A}^2\right)\left(m^2 + \hat{p}^2\right)^{-1}} \left(m^2 + \hat{p}^2\right)^{\frac{1}{2}}. \tag{7}$$

With the above-mentioned assumption, it is now very tempting to expand the first square root factor. A very similar approach has already been proposed by [3]. We would like to restrict ourselves to a series expansion of the kind

$$\sqrt{1+\hat{y}} \approx 1 + \tfrac{1}{2}\hat{y} - \tfrac{1}{8}\hat{y}^2 + \ldots \tag{8}$$

containing only constant, linear and quadratic terms, where $\hat{y}$ denotes

$$\hat{y} = \left(-e\hat{\vec{p}}\cdot\vec{A} - e\vec{A}\cdot\hat{\vec{p}} + e^2\vec{A}^2\right)\left(m^2 + \hat{p}^2\right)^{-1}. \tag{9}$$

Hamiltonian (7) is therefore approximated by

$$\hat{H}' \approx \hat{H}'_0 + \hat{H}'_1 + \hat{H}'_2 \tag{10}$$

with

$$\hat{H}'_1 = e\left[-\tfrac{1}{2}\left(\hat{\vec{p}}\cdot\vec{A} + \vec{A}\cdot\hat{\vec{p}}\right)\left(m^2 + \hat{p}^2\right)^{-\frac{1}{2}} + A^0\right], \tag{11}$$

$$\hat{H}'_2 = e^2\left[\tfrac{1}{2}\vec{A}^2\left(m^2 + \hat{p}^2\right)^{-\frac{1}{2}} - \tfrac{1}{8}\left(\hat{\vec{p}}\cdot\vec{A} + \vec{A}\cdot\hat{\vec{p}}\right)\left(m^2 + \hat{p}^2\right)^{-1}\left(\hat{\vec{p}}\cdot\vec{A} + \vec{A}\cdot\hat{\vec{p}}\right)\left(m^2 + \hat{p}^2\right)^{-\frac{1}{2}}\right], \tag{12}$$

where we have reordered the terms of expansion (9), retaining only terms to (and including the) quadratic order in *e* and recollected powers of $\left(m^2 + \hat{p}^2\right)$. What we have won by (10) is a separation of the free Hamiltonian (6) from the coupling terms in (5) that approximately result in the sum of $\hat{H}'_1$ and $\hat{H}'_2$ (i.e. (11) and (12) respectively). For we are only interested in corrections to the free Hamiltonian anyway, this approximation might not hurt very much. But however, this separation seems not to be a true one, because of the multiple factors of powers of $\left(m^2 + \hat{p}^2\right)$ in (11) and (12). That is, we need an interpretation of these operators. To this end, it is useful to know that for the free square root operator equation (2), an integral representation can be given (see [1,2]):

$$i\partial_t \phi(\vec{x},t) = \int d^3x' \, \Omega(\vec{x}-\vec{x}')\phi(\vec{x}',t) =: (\Omega\phi)(\vec{x},t), \tag{13}$$

where $\Omega$ denotes an energy distribution

$$\Omega(\vec{x}-\vec{x}') = \int \frac{d^3p}{(2\pi)^3} \omega_p \, e^{-i\vec{p}\cdot(\vec{x}-\vec{x}')} \tag{14}$$

with

$$\omega_p = \omega_{\vec{p}} = \sqrt{m^2 + \vec{p}^2} \ . \tag{15}$$

(13) results from the fact, that one would expect to obtain the following momentum space representation of (2):

$$i\partial_t \tilde{\phi}(\vec{p},t) = \omega_p \tilde{\phi}(\vec{p},t)$$

with $\tilde{\phi}(\vec{p},t)$ denoting the Fourier-transformed $\phi(\vec{x},t)$. If the operator $\sqrt{m^2 + \hat{\vec{p}}^2}$ corresponds to $\int d^3 x' \Omega(\vec{x} - \vec{x}')$, the operator $\left(m^2 + \hat{\vec{p}}^2\right)^{-\frac{1}{2}}$ must correspond to $\int d^3 x' \Omega^{-1}(\vec{x} - \vec{x}')$ with

$$\Omega^{-1}(\vec{x} - \vec{x}') = \int \frac{d^3 p}{(2\pi)^3} \omega_p^{-1} e^{-i\vec{p}\cdot(\vec{x}-\vec{x}')} \ , \tag{16}$$

because

$$\int d^3 x'' \Omega^{-1}(\vec{x} - \vec{x}'') \Omega(\vec{x}'' - \vec{x}') = \int d^3 x'' \Omega(\vec{x} - \vec{x}'') \Omega^{-1}(\vec{x}'' - \vec{x}') = \delta^3(\vec{x} - \vec{x}') \tag{17}$$

with the Dirac distribution

$$\delta^3(\vec{x} - \vec{x}') = \int \frac{d^3 p}{(2\pi)^3} e^{-i\vec{p}\cdot(\vec{x}-\vec{x}')} \ . \tag{18}$$

(17) should be an integral representation of the "symbolic equation"

$$\left(m^2 + \hat{\vec{p}}^2\right)^{-\frac{1}{2}} \left(m^2 + \hat{\vec{p}}^2\right)^{\frac{1}{2}} = \left(m^2 + \hat{\vec{p}}^2\right)^{\frac{1}{2}} \left(m^2 + \hat{\vec{p}}^2\right)^{-\frac{1}{2}} = 1 \ .$$

Accordingly, terms with the *n*th power of $\left(m^2 + \hat{\vec{p}}^2\right)^{\frac{1}{2}}$,

$$\left(m^2 + \hat{\vec{p}}^2\right)^{\frac{n}{2}} \ , \tag{19}$$

correspond to integrals over "the *n*th power of $\Omega$":

$$\Omega^n(\vec{x} - \vec{x}') = \int \frac{d^3 p}{(2\pi)^3} \omega_p^n e^{-i\vec{p}\cdot(\vec{x}-\vec{x}')} \ . \tag{20}$$

By replacing the operators of type (19) by integrals over "powers of $\Omega$" as given in (20), $\hat{H}'_1$ and $\hat{H}'_2$ (see (11) and (12), respectively) can now be given a configuration space representation, too.
With these preparations, we can now address to the quantisation of the scalar field with the aim to be able to calculate scattering matrix elements.

**Quantisation of the scalar field and the description of scattering processes**

Starting with Hamiltonian (10), it is now possible to describe scattering processes within the framework of quantum field theory. For free scalar particles, a quantum field theoretic ansatz is described e.g. in [2] and [4], using (2) and (13), respectively, as equations for a field operator $\phi(x)$.

The latter one can be formulated with the help of creation and annihilation operators $\hat{a}_{\vec{p}}^+$ and $\hat{a}_{\vec{p}}$, respectively:

$$\phi(x) = \int \frac{d^3 p}{(2\pi)^{3/2}} e^{-ip\cdot x} \hat{a}_{\vec{p}}, \tag{21}$$

where as usual $p \cdot x = p_\mu x^\mu = \omega_p t - \vec{p} \cdot \vec{x}$ with the 4-vector $p = (\omega_p, \vec{p})$ and the subsequent definitions are postulated:

$$\hat{a}_{\vec{p}} |0\rangle = 0, \tag{22}$$

$$\langle 0 | \hat{a}_{\vec{p}}^+ = 0, \tag{23}$$

$$[\hat{a}_{\vec{p}}, \hat{a}_{\vec{p}'}^+] = \delta^3(\vec{p} - \vec{p}'), \tag{24 a}$$

$$[\hat{a}_{\vec{p}}, \hat{a}_{\vec{p}'}] = 0, \quad [\hat{a}_{\vec{p}}^+, \hat{a}_{\vec{p}'}^+] = 0 \tag{24 b}$$

with the vacuum state $|0\rangle$. Since we are interested in a quantum theory for bosons, $[\bullet, \bullet]$ in (24) must be a commutator (for fermions we would use here an anti-commutator instead, cf. e.g. [4]). Equations (21) to (24) are identical to those that one would postulate within a non-relativistic quantum field theory for bosons.

For the density of a Hamiltonian, we make the usual ansatz (see e.g. [7]):

$$\hat{H} = \phi^+(x) \hat{H}' \phi(x) \tag{25}$$

which one can retrieve from a density of a Lagrangian (see [2]):

$$L = \tfrac{i}{2}\left(\phi^+ \partial_t \phi - \phi \partial_t \phi^+\right) - \tfrac{1}{2}\phi^+(\Omega\phi) - \tfrac{1}{2}(\Omega\phi)^+ \phi. \tag{25 a}$$

Substituting (10) into (25), we get

$$\hat{H} \approx \hat{H}_0 + \hat{H}_1 + \hat{H}_2 \tag{26}$$

with

$$\hat{H}_0 = \phi^+(x) \hat{H}_0' \phi(x), \tag{27}$$

$$\hat{H}_1 = \phi^+(x) \hat{H}_1' \phi(x), \tag{28}$$

$$\hat{H}_2 = \phi^+(x) \hat{H}_2' \phi(x). \tag{29}$$

(25) is (among other things) motivated by the fact that

$$\int d^3 x \, \hat{H}_0 = \int d^3 p \, \omega_p \, \hat{a}_{\vec{p}}^+ \hat{a}_{\vec{p}} \tag{30}$$

reproduces the relativistic analogue of the free non-relativistic Hamiltonian:

$$\int d^3 p \, \frac{\vec{p}^{\,2}}{2m} \hat{a}_{\vec{p}}^+ \hat{a}_{\vec{p}}. \tag{31}$$

(28) together with (29) are the densities of the Hamiltonian to (and including the) quadratic order in *e*. (21) to (24) are valid for free particles, but can also be used for interacting ones, if Dirac's representation is used instead of the so far applied Heisenberg representation. Then, with (28) and (29) combined to a Hamiltonian density

$$\hat{H}^I = \hat{H}_1 + \hat{H}_2 \tag{32}$$

for the interaction of scalar bosons with photons, we can now start to calculate scattering matrix elements. To this purpose, we need the serial expansion of the *S*-operator (see e.g. [7]) to the order of $e^2$:

$$\hat{S} = 1 + \hat{S}^{(1)} + \hat{S}^{(2)} + \ldots \tag{33}$$

with

$$\hat{S}^{(1)} = -i \int d^4 x\, T\!\left(\hat{H}^I(x)\right), \tag{34}$$

$$\hat{S}^{(2)} = \tfrac{1}{2}(-i)^2 \int d^4 x_1 \int d^4 x_2\, T\!\left(\hat{H}^I(x_1)\hat{H}^I(x_2)\right), \tag{35}$$

where we have introduced a time ordering operator

$$T\!\left(\hat{H}^I(x_1)\hat{H}^I(x_2)\right) = \theta(x_1^0 - x_2^0)\hat{H}^I(x_1)\hat{H}^I(x_2) + \theta(x_2^0 - x_1^0)\hat{H}^I(x_2)\hat{H}^I(x_1) = T\!\left(\hat{H}^I(x_2)\hat{H}^I(x_1)\right) \tag{36}$$

with

$$\theta(t) = \begin{cases} 1, t \geq 0 \\ 0, t < 0 \end{cases}. \tag{37}$$

$\hat{S}^{(1)}$ does not only contribute to the expansion (33) with terms of order *e*, but also to order $e^2$. Therefore, we can split off $\hat{S}^{(1)}$ into a term $\hat{S}_1$ containing only terms in *e*,

$$\hat{S}_1 = (-ie)\int d^4 x\, T\!\left(-\tfrac{1}{2}\phi^+(x)\!\left(\hat{\vec{p}}\cdot\vec{A}(x) + \vec{A}(x)\cdot\hat{\vec{p}}\right)\!\left(\Omega^{-1}\phi\right)\!(x) + \phi^+(x)A^0(x)\phi(x)\right) \tag{38}$$

with

$$\left(\Omega^{-1}\phi\right)\!(x) = \int d^3 x_1\, \Omega^{-1}(\vec{x} - \vec{x}_1)\phi(\vec{x}_1, t) \tag{39}$$

and a part containing only terms in $e^2$:

$$\begin{aligned}\hat{S}_{12} = (-ie^2)\int d^4 x_1\, T\Big(&\tfrac{1}{2}\phi^+(x_1)\vec{A}^2(x_1)\!\left(\Omega^{-1}\phi\right)\!(x_1) \\ -&\tfrac{1}{8}\phi^+(x_1)\!\left(\hat{\vec{p}}_1\cdot\vec{A}(x_1) + \vec{A}(x_1)\cdot\hat{\vec{p}}_1\right)\!\int d^3 x_2\, \Omega^{-2}(\vec{x}_1 - \vec{x}_2)\!\left(\hat{\vec{p}}_2\cdot\vec{A}(\vec{x}_2,t_1) + \vec{A}(\vec{x}_2,t_1)\cdot\hat{\vec{p}}_2\right)\!\left(\Omega^{-1}\phi\right)\!(\vec{x}_2,t_1)\Big)\end{aligned} \tag{40}$$

where the momentum operator $\hat{\vec{p}}_1$ contains a gradient acting on $\vec{x}_1$ and $\hat{\vec{p}}_2$ acting on $\vec{x}_2$. In (38) and (40), we have already substituted (32) into (34) and replaced powers of $\left(m^2 + \hat{\vec{p}}^2\right)^{\frac{1}{2}}$ by integrals over "powers of $\Omega$" (see (20)) in (11). Thus we can rewrite (34) as

$$\hat{S}^{(1)} = \hat{S}_1 + \hat{S}_{12}. \tag{41}$$

The time ordering operator appearing in $\hat{S}^{(1)}$ can be left out, because it contains only one time. In $\hat{S}^{(2)}$ we only want to retain terms of order $e^2$, therefore $\hat{H}^I$ can be approximated by $\hat{H}_1$:

$$\hat{S}^{(2)} = (-ie)^2 \tfrac{1}{2} \int d^4x_1 \int d^4x_2 \, T\Big(-\tfrac{1}{2}\phi^+(x_1)\big(\hat{\vec{p}}_1 \cdot \vec{A}(x_1) + \vec{A}(x_1)\cdot \hat{\vec{p}}_1\big)\big(\Omega^{-1}\phi\big)(x_1)\phi^+(x_2)A^0(x_2)\phi(x_2)$$
$$+ \tfrac{1}{4}\phi^+(x_1)\big(\hat{\vec{p}}_1\cdot\vec{A}(x_1)+\vec{A}(x_1)\cdot\hat{\vec{p}}_1\big)\big(\Omega^{-1}\phi\big)(x_1)\phi^+(x_2)\big(\hat{\vec{p}}_2\cdot\vec{A}(x_2)+\vec{A}(x_2)\cdot\hat{\vec{p}}_2\big)\big(\Omega^{-1}\phi\big)(x_2) \quad (42)$$
$$+ \phi^+(x_1)A^0(x_1)\phi(x_1)\phi^+(x_2)A^0(x_2)\phi(x_2)\Big)$$

Here, the time ordering operator must be taken into account, because there are two times: $t_1$ and $t_2$. We have already quantised the field of scalar bosons, but must do now the corresponding with the electromagnetic field. Since the scalar field in (38), (40) and (42) couple in a different way to the vector potential $\vec{A}$ than to the scalar potential $A^0$, the choice of a Coulomb gauge seems to be appropriate:

$$\vec{\nabla}\cdot\vec{A} = 0 . \qquad (43)$$

Then the field equations take on the form

$$\big(\partial_t^2 - \vec{\nabla}^2\big)\vec{A} + \partial_t \vec{\nabla} A^0 = \vec{j} , \qquad (44\text{ a})$$
$$\vec{\nabla}^2 A^0 = -\rho \qquad (44\text{ b})$$

with charge and current densities $\rho$ and $\vec{j}$, respectively. From (44 b) we can see that in this gauge the scalar potential is just a c-number:

$$A^0(\vec{x},t) = \frac{1}{4\pi}\int d^3x' \frac{\rho(\vec{x}',t)}{|\vec{x}-\vec{x}'|} , \qquad (45)$$

whereas the vector potential $\vec{A}$ becomes an operator when being quantised. For a free field, $\vec{A}$ can be chosen like (see e.g. [4,7,10])

$$\hat{\vec{A}}(x) = \int \frac{d^3k}{\sqrt{(2\pi)^3 2\tilde{\omega}_k}} \sum_{\lambda=1}^{2} \big(\hat{c}_{\vec{k}\lambda} e^{-ik\cdot x} + \hat{c}^+_{\vec{k}\lambda} e^{ik\cdot x}\big)\vec{\varepsilon}(\vec{k},\lambda) \qquad (46)$$

with the usual photon frequency

$$\tilde{\omega}_k = \sqrt{\vec{k}^2}$$

and with creation and annihilation operators $\hat{c}_{\vec{k}\lambda}$ and $\hat{c}^+_{\vec{k}\lambda}$, respectively, for photons:

$$\big[\hat{c}_{\vec{k}'\lambda'}, \hat{c}^+_{\vec{k}\lambda}\big] = \delta_{\lambda'\lambda}\,\delta^3(\vec{k}'-\vec{k}),\; \big[\hat{c}_{\vec{k}'\lambda'}, \hat{c}_{\vec{k}\lambda}\big] = \big[\hat{c}^+_{\vec{k}'\lambda'}, \hat{c}^+_{\vec{k}\lambda}\big] = 0 , \qquad (47)$$

and $A^0$ would even vanish. The polarisation vectors $\vec{\varepsilon}(\vec{k},\lambda)$ fulfil the relation (see e.g. [4, 7]):

$$\sum_{\lambda=1}^{2}\varepsilon^i(\vec{k},\lambda)\varepsilon^j(\vec{k},\lambda) = \delta_{ij} - \frac{k_i k_j}{\vec{k}^2} . \qquad (48)$$

Due to the Coulomb gauge condition (43),

$$\vec{k} \cdot \vec{\varepsilon}(\vec{k}, \lambda) = 0 \qquad (49)$$

is valid, too.

In the following sections, we are going to calculate $\hat{S}^{(1)}$ and $\hat{S}^{(2)}$ by substituting the field operators $\phi(x)$ and $\vec{A}(x)$ from (21) and (46) as well as the distributions (20). Since we are considering electromagnetic interactions between (charged) spin-0 bosons, we have to take $A^0$ in (45) into account, too. Therefore, we first have to find out what the density of charge in (45) will be in this case. This can be done by coupling the Lagrangian density for free spin-0 bosons (25 a) to an electromagnetic field with the aid of the minimal coupling scheme $i\partial_\mu \to i\partial_\mu - eA_\mu$. Then in that Lagragian density an extra term $eA^0 \phi^+ \phi$ (the only one with an $A^0$) emerges. If one subsequently regards the sum of that spin-0 and the electromagnetic Lagrangian, it is possible to obtain from it the electromagnetic field equations by means of the Euler-Lagrange equations. The former equations then contain a charge density $e\phi^+\phi$. That is why we can set $\rho$ in (45) to $\phi^+\phi$.

With these results, one can then continue to calculate scattering matrix elements.

The first term we calculate is

$$\hat{I}_1 := \int d^4x \, \phi^+(x) \left( \hat{\vec{p}} \cdot \vec{A}(x) + \vec{A}(x) \cdot \hat{\vec{p}} \right) \left( \Omega^{-1} \phi \right)(x) \qquad (50)$$

appearing in $\hat{S}^{(1)}$ (see (38)). To this end, it is useful to recognise that by means of (16) we get

$$\left( \Omega^{-1} \phi \right)(x) = \int \frac{d^3p}{(2\pi)^{3/2}} \, \omega_p^{-1} \, e^{-ip \cdot x} \, \hat{a}_{\vec{p}} \, . \qquad (51)$$

With (51) and integration by parts, (50) yields:

$$\hat{I}_1 = \int \frac{d^3k \, d^3p_1 \, d^3p_2}{\sqrt{2\pi} \sqrt{2\tilde{\omega}_k}} \sum_\lambda \vec{\varepsilon}(\vec{k},\lambda) \cdot \frac{(\vec{p}_1 + \vec{p}_2)}{\omega_{p_1}} \hat{a}^+_{\vec{p}_2} \hat{a}_{\vec{p}_1} \left( \delta^4(p_1 - p_2 + k) \hat{c}_{\vec{k}\lambda} + \delta^4(p_1 - p_2 - k) \hat{c}^+_{\vec{k}\lambda} \right), \qquad (52)$$

where

$$\delta^4(p_1 - p_2 \pm k) = \delta(\omega_{p_1} - \omega_{p_2} \pm \omega_k) \delta^3(\vec{p}_1 - \vec{p}_2 \pm \vec{k}) \qquad (53)$$

and, of course, the $a$ operators commute with the $c$ operators. In $\hat{S}_{12}$ (40), the first integral can be expressed in a similar way:

$$\hat{I}_2 := \int d^4x \, \phi^+(x) \vec{A}^2(x) \left( \Omega^{-1} \phi \right)(x) =$$
$$\int \frac{d^3k_1 \, d^3k_2 \, d^3p_1 \, d^3p_2}{\sqrt{2\pi} \sqrt{2\tilde{\omega}_{k_1} 2\tilde{\omega}_{k_2}}} \frac{1}{\omega_{p_2}} \hat{a}^+_{\vec{p}_1} \hat{a}_{\vec{p}_2} \sum_{\lambda_1,\lambda_2} \vec{\varepsilon}(\vec{k}_1,\lambda_1) \cdot \vec{\varepsilon}(\vec{k}_2,\lambda_2) \cdot$$
$$\left( \delta^4(k_1 + k_2 + p_1 - p_2) \hat{c}^+_{\vec{k}_1\lambda_1} \hat{c}^+_{\vec{k}_2\lambda_2} + \delta^4(k_1 - k_2 + p_1 - p_2) \hat{c}^+_{\vec{k}_1\lambda_1} \hat{c}_{\vec{k}_2\lambda_2} + \right.$$
$$\left. \delta^4(-k_1 + k_2 + p_1 - p_2) \hat{c}_{\vec{k}_1\lambda_1} \hat{c}^+_{\vec{k}_2\lambda_2} + \delta^4(-k_1 - k_2 + p_1 - p_2) \hat{c}_{\vec{k}_1\lambda_1} \hat{c}_{\vec{k}_2\lambda_2} \right) \qquad (54)$$

After a quite lengthy but straightforward calculation, the second integral in (40) yields

$$\hat{I}_3 := \int d^4x_1 \int d^3x_2 \, \phi^+(x_1)\left(\hat{\vec{p}}_1 \cdot \vec{A}(x_1) + \vec{A}(x_1)\cdot \hat{\vec{p}}_1\right)\Omega^{-2}(\vec{x}_1 - \vec{x}_2)\cdot \qquad (55)$$
$$\left(\hat{\vec{p}}_2 \cdot \vec{A}(\vec{x}_2,t_1) + \vec{A}(\vec{x}_2,t_1)\cdot \hat{\vec{p}}_2\right)\left(\Omega^{-1}\phi\right)(\vec{x}_2,t_1) =$$

$$\int \frac{d^3k_1 \, d^3k_2 \, d^3p_1 \, d^3p_2}{(2\pi)^2 \sqrt{2\tilde{\omega}_{k_1} 2\tilde{\omega}_{k_2}}} \frac{1}{\omega_{p_1}} \hat{a}^+_{\vec{p}_2} \hat{a}_{\vec{p}_1} \sum_{\lambda_2} \vec{\varepsilon}(\vec{k}_2,\lambda_2)\cdot (\vec{p}_1 + \vec{p}_2)\cdot$$

$$\left(\sum_{\lambda_1} \vec{\varepsilon}(\vec{k}_1,\lambda_1)\cdot \frac{(2\vec{p}_1 + \vec{k}_1)}{\omega^2_{\vec{p}_1+\vec{k}_1}} \left(\delta^4(-k_1 + k_2 + p_1 - p_2)\hat{c}_{\vec{k}_2\lambda_2}\hat{c}_{\vec{k}_1\lambda_1} + \delta^4(k_1 - k_2 + p_1 - p_2)\hat{c}^+_{\vec{k}_2\lambda_2}\hat{c}_{\vec{k}_1\lambda_1}\right) + \right.$$

$$\left.\sum_{\lambda_1} \vec{\varepsilon}(\vec{k}_1,\lambda_1)\cdot \frac{(2\vec{p}_1 - \vec{k}_1)}{\omega^2_{\vec{p}_1-\vec{k}_1}} \left(\delta^4(-k_1 + k_2 + p_1 - p_2)\hat{c}_{\vec{k}_2\lambda_2}\hat{c}^+_{\vec{k}_1\lambda_1} + \delta^4(-k_1 - k_2 + p_1 - p_2)\hat{c}^+_{\vec{k}_2\lambda_2}\hat{c}^+_{\vec{k}_1\lambda_1}\right)\right)$$

Finally, we want to determine the second term in $\hat{S}^{(2)}$ (see (42)) which is not just like the product of two operators $\hat{I}_1$ due to the time ordering operator as defined in (36). Unfortunately, we cannot use the famous Wick theorem, because the scalar field operator (21) contains only contributions to positive energy solutions: it does not consist of a sum of both positive and negative energy solutions as it would be the case for the field operator of the Klein-Gordon equation. Due to the symmetry of the time ordering operator (36) in its arguments, we may conclude

$$\int d^4x_1 \int d^4x_2 \, T\!\left(\hat{H}^I(x_1)\hat{H}^I(x_2)\right) = 2\int d^4x_1 \int d^4x_2 \, \theta(x_1^0 - x_2^0)\hat{H}^I(x_1)\hat{H}^I(x_2). \qquad (56)$$

With this property, the calculation of the second term in (42) can be simplified a bit:

$$\hat{I}_4 := \int d^4x_1 \int d^4x_2 \, T\!\left(\phi^+(x_1)\!\left(\hat{\vec{p}}_1\cdot\vec{A}(x_1)+\vec{A}(x_1)\cdot\hat{\vec{p}}_1\right)\!\left(\Omega^{-1}\phi\right)(x_1)\cdot \right. \qquad (57\text{ a})$$
$$\left. \phi^+(x_2)\!\left(\hat{\vec{p}}_2\cdot\vec{A}(x_2)+\vec{A}(x_2)\cdot\hat{\vec{p}}_2\right)\!\left(\Omega^{-1}\phi\right)(x_2)\right) =$$

$$2\int \frac{d^3k' \, d^3k \, d^3p'_1 \, d^3p'_2 \, d^3p_1 \, d^3p_2}{2\pi \sqrt{2\tilde{\omega}_{k'} 2\tilde{\omega}_k}} \hat{a}^+_{\vec{p}'_2}\hat{a}_{\vec{p}'_1}\hat{a}^+_{\vec{p}_2}\hat{a}_{\vec{p}_1} \sum_{\lambda'} \vec{\varepsilon}(\vec{k}',\lambda')\cdot \frac{(\vec{p}'_1+\vec{p}'_2)}{\omega_{p'_1}} \sum_{\lambda} \vec{\varepsilon}(\vec{k},\lambda)\cdot \frac{(\vec{p}_1+\vec{p}_2)}{\omega_{p_1}}\cdot$$

$$\left(\hat{c}_{\vec{k}'\lambda'}\hat{c}_{\vec{k}\lambda}\,\delta^3(\vec{p}'_1-\vec{p}'_2+\vec{k}')\delta^3(\vec{p}_1-\vec{p}_2+\vec{k})\cdot\right.$$
$$\int \frac{dt_1 dt_2}{(2\pi)^2}\theta(t_1-t_2)\exp\!\left(-i(\omega_{p'_1}-\omega_{p'_2}+\tilde{\omega}_{k'})t_1\right)\exp\!\left(-i(\omega_{p_1}-\omega_{p_2}+\tilde{\omega}_k)t_2\right)+$$

$$\hat{c}_{\vec{k}'\lambda'}\hat{c}^+_{\vec{k}\lambda}\,\delta^3(\vec{p}'_1-\vec{p}'_2+\vec{k}')\delta^3(\vec{p}_1-\vec{p}_2-\vec{k})\cdot$$
$$\int \frac{dt_1 dt_2}{(2\pi)^2}\theta(t_1-t_2)\exp\!\left(-i(\omega_{p'_1}-\omega_{p'_2}+\tilde{\omega}_{k'})t_1\right)\exp\!\left(-i(\omega_{p_1}-\omega_{p_2}-\tilde{\omega}_k)t_2\right)+$$

$$\hat{c}^+_{\vec{k}'\lambda'}\hat{c}_{\vec{k}\lambda}\,\delta^3(\vec{p}'_1-\vec{p}'_2-\vec{k}')\delta^3(\vec{p}_1-\vec{p}_2+\vec{k})\cdot$$
$$\int \frac{dt_1 dt_2}{(2\pi)^2}\theta(t_1-t_2)\exp\!\left(-i(\omega_{p'_1}-\omega_{p'_2}-\tilde{\omega}_{k'})t_1\right)\exp\!\left(-i(\omega_{p_1}-\omega_{p_2}+\tilde{\omega}_k)t_2\right)+$$

$$\hat{c}^+_{\vec{k}'\lambda'}\hat{c}^+_{\vec{k}\lambda}\,\delta^3(\vec{p}'_1-\vec{p}'_2-\vec{k}')\delta^3(\vec{p}_1-\vec{p}_2-\vec{k})\cdot$$
$$\left.\int \frac{dt_1 dt_2}{(2\pi)^2}\theta(t_1-t_2)\exp\!\left(-i(\omega_{p'_1}-\omega_{p'_2}-\tilde{\omega}_{k'})t_1\right)\exp\!\left(-i(\omega_{p_1}-\omega_{p_2}-\tilde{\omega}_k)t_2\right)\right)$$

The two integrals over the $\theta$ function can be performed by means of the introduction of the two variables

$$\tau := t_1 - t_2,$$
$$T := t_1 + t_2 \tag{58}$$

with the Jacobian

$$\frac{\partial(t_1, t_2)}{\partial(\tau, T)} = \frac{1}{2}.$$

A linear combination of these variables

$$A\tau + BT = a\, t_1 + b\, t_2 \tag{59 a}$$

can be expressed by means of

$$A = \tfrac{1}{2}(a - b), \tag{59 b}$$
$$B = \tfrac{1}{2}(a + b).$$

(57) contains four terms of the subsequent type that can be simplified with the help of (58) and (59):

$$\int \frac{dt_1\, dt_2}{(2\pi)^2} \theta(t_1 - t_2) \exp(-i\omega_1 t_1) \exp(-i\omega_2 t_2) =$$
$$\tfrac{1}{2} \int \frac{d\tau}{2\pi} \theta(\tau) \exp\!\left(-\tfrac{i}{2}(\omega_1 - \omega_2)\tau\right) \int \frac{dT}{2\pi} \exp\!\left(-\tfrac{i}{2}(\omega_1 + \omega_2)T\right) = \tag{60 a}$$
$$\delta(\omega_1 + \omega_2) \int \frac{d\tau}{2\pi} \theta(\tau) \exp\!\left(-\tfrac{i}{2}(\omega_1 - \omega_2)\tau\right)$$

The function $\theta$ can be expressed by an integral in the complex plane (see e.g. [6,8]),

$$\theta(\pm\tau) = \pm i \int_{-\infty}^{\infty} \frac{dp_0}{2\pi} e^{-ip_0 \tau} \frac{1}{p_0 \pm i\varepsilon} \tag{61}$$

with an $\varepsilon$ approaching zero. Substituting this into (60 a), we get:

$$\delta(\omega_1 + \omega_2) \int \frac{d\tau}{2\pi} \theta(\tau) \exp\!\left(-\tfrac{i}{2}(\omega_1 - \omega_2)\tau\right) = \frac{i}{2\pi} \frac{1}{\omega_2 + i\varepsilon} \delta(\omega_1 + \omega_2). \tag{60 b}$$

With this result, (57) becomes:

$$\hat{I}_4 = 2i \int \frac{d^3k' d^3k \, d^3p_1' \, d^3p_2' d^3p_1 \, d^3p_2}{(2\pi)^2 \sqrt{2\tilde{\omega}_{k'} 2\tilde{\omega}_k}} \, \hat{a}_{\vec{p}_2'}^+ \hat{a}_{\vec{p}_1'} \hat{a}_{\vec{p}_2}^+ \hat{a}_{\vec{p}_1} \sum_{\lambda'} \vec{\varepsilon}(\vec{k}', \lambda') \cdot \frac{(\vec{p}_1' + \vec{p}_2')}{\omega_{p_1'}} \sum_{\lambda} \vec{\varepsilon}(\vec{k}, \lambda) \cdot \frac{(\vec{p}_1 + \vec{p}_2)}{\omega_{p_1}} \cdot$$

$$\left( \hat{c}_{\vec{k}'\lambda'} \, \hat{c}_{\vec{k}\lambda} \, \delta(\omega_{p_1'} - \omega_{p_2'} + \tilde{\omega}_{k'} + \omega_{p_1} - \omega_{p_2} + \tilde{\omega}_k) \delta^3(\vec{p}_1' - \vec{p}_2' + \vec{k}') \delta^3(\vec{p}_1 - \vec{p}_2 + \vec{k}) \cdot \frac{1}{\omega_{p_1} - \omega_{p_2} + \tilde{\omega}_k + i\varepsilon} + \right.$$

$$\hat{c}_{\vec{k}'\lambda'} \, \hat{c}_{\vec{k}\lambda}^+ \, \delta(\omega_{p_1'} - \omega_{p_2'} + \tilde{\omega}_{k'} + \omega_{p_1} - \omega_{p_2} - \tilde{\omega}_k) \delta^3(\vec{p}_1' - \vec{p}_2' + \vec{k}') \delta^3(\vec{p}_1 - \vec{p}_2 - \vec{k}) \cdot \frac{1}{\omega_{p_1} - \omega_{p_2} - \tilde{\omega}_k + i\varepsilon} +$$

$$\hat{c}_{\vec{k}'\lambda'}^+ \, \hat{c}_{\vec{k}\lambda} \, \delta(\omega_{p_1'} - \omega_{p_2'} - \tilde{\omega}_{k'} + \omega_{p_1} - \omega_{p_2} + \tilde{\omega}_k) \delta^3(\vec{p}_1' - \vec{p}_2' - \vec{k}') \delta^3(\vec{p}_1 - \vec{p}_2 + \vec{k}) \cdot \frac{1}{\omega_{p_1} - \omega_{p_2} + \tilde{\omega}_k + i\varepsilon} +$$

$$\left. \hat{c}_{\vec{k}'\lambda'}^+ \hat{c}_{\vec{k}\lambda}^+ \, \delta(\omega_{p_1'} - \omega_{p_2'} - \tilde{\omega}_{k'} + \omega_{p_1} - \omega_{p_2} - \tilde{\omega}_k) \delta^3(\vec{p}_1' - \vec{p}_2' - \vec{k}') \delta^3(\vec{p}_1 - \vec{p}_2 - \vec{k}) \cdot \frac{1}{\omega_{p_1} - \omega_{p_2} - \tilde{\omega}_k + i\varepsilon} \right)$$

(57 b)

So far, we have only considered terms of the scattering matrix containing the electromagnetic vector potential $\vec{A}$. Now, we have to address to those terms containing the scalar potential $A^0$ too. (38) does not only contain (50), but also a Coulomb potential term:

$$\hat{I}_5 = \int d^4x \, \phi^+ A^0 \phi \,. \tag{62 a}$$

By substituting $\rho$ in (45) by $\phi^+\phi$, we obtain the following equation for (62 a), if we take into account that the Fourier transformed Coulomb potential looks like

$$\int d^3x \frac{e^{-i\vec{k}\cdot\vec{x}}}{|\vec{x}|} = -\frac{4\pi}{\vec{k}^2} : \tag{63}$$

$$\hat{I}_5 = -\frac{1}{4(2\pi)^2} \int d^3p' \, d^3p \, d^3k' \, d^3k \, \delta^4(p' + k' - p - k) \hat{a}_{\vec{p}'}^+ \hat{a}_{\vec{k}'}^+ \hat{a}_{\vec{k}} \hat{a}_{\vec{p}} \frac{1}{(\vec{k}' - \vec{k})^2} \,. \tag{62 b}$$

(42) contains two terms in $A^0$. The first term consists of a combination of (50) and (62 a), but taken at different times and therefore joined via the time ordering operator. That is why we have to use (60) as well as (45) and (63) again:

$$\hat{I}_6 = \int d^4x_1 d^4x_2 \, T\left(\phi^+(x_1)(\hat{\vec{p}}_1 \cdot \vec{A}(x_1) + \vec{A}(x_1) \cdot \hat{\vec{p}}_1)(\Omega^{-1}\phi)(x_1)\phi^+(x_2) A^0(x_2)\phi(x_2)\right) =$$

$$\frac{i}{2^{3/2}(2\pi)^7} \int d^3k_1 \, d^3p_2 \, d^3p_1 \, d^3p' \, d^3p \, d^3k' \, d^3k \, \frac{1}{\sqrt{\tilde{\omega}_{k_1}} \, \omega_{p_1}} \sum_{\lambda=1}^{2} \vec{\varepsilon}(\vec{k}_1, \lambda) \cdot (\vec{p}_1 + \vec{p}_2) \frac{\delta^3(\vec{p}' + \vec{k}' - \vec{p} - \vec{k})}{(\vec{k}' - \vec{k})^2} \cdot$$

$$\left[ \delta^3(\vec{p}_1 - \vec{p}_2 + \vec{k}_1) \delta(\omega_k + \omega_p - \omega_{k'} - \omega_{p'} + \omega_{p_1} - \omega_{p_2} + \tilde{\omega}_{k_1}) \frac{1}{\omega_k + \omega_p - \omega_{k'} - \omega_{p'} + i\varepsilon} \, \hat{a}_{\vec{p}_1} \hat{a}_{\vec{p}'}^+ \hat{a}_{\vec{k}'}^+ \hat{a}_{\vec{k}} \hat{a}_{\vec{p}} \hat{c}_{\vec{k}_1 \lambda} \right.$$

$$+ \delta^3(\vec{p}_1 - \vec{p}_2 + \vec{k}_1) \delta(\omega_k + \omega_p - \omega_{k'} - \omega_{p'} + \omega_{p_1} - \omega_{p_2} + \tilde{\omega}_{k_1}) \frac{1}{\omega_{p_1} - \omega_{p_2} + \tilde{\omega}_{k_1} + i\varepsilon} \, \hat{a}_{\vec{p}'}^+ \hat{a}_{\vec{k}'}^+ \hat{a}_{\vec{k}} \hat{a}_{\vec{p}} \hat{a}_{\vec{p}_1} \hat{c}_{\vec{k}_1 \lambda}$$

$$+ \delta^3(\vec{p}_1 - \vec{p}_2 - \vec{k}_1) \delta(\omega_k + \omega_p - \omega_{k'} - \omega_{p'} + \omega_{p_1} - \omega_{p_2} - \tilde{\omega}_{k_1}) \frac{1}{\omega_k + \omega_p - \omega_{k'} - \omega_{p'} + i\varepsilon} \, \hat{a}_{\vec{p}_1} \hat{a}_{\vec{p}'}^+ \hat{a}_{\vec{k}'}^+ \hat{a}_{\vec{k}} \hat{a}_{\vec{p}} \hat{c}_{\vec{k}_1 \lambda}^+$$

$$\left. + \delta^3(\vec{p}_1 - \vec{p}_2 - \vec{k}_1) \delta(\omega_k + \omega_p - \omega_{k'} - \omega_{p'} + \omega_{p_1} - \omega_{p_2} - \tilde{\omega}_{k_1}) \frac{1}{\omega_{p_1} - \omega_{p_2} - \tilde{\omega}_{k_1} + i\varepsilon} \, \hat{a}_{\vec{p}'}^+ \hat{a}_{\vec{k}'}^+ \hat{a}_{\vec{k}} \hat{a}_{\vec{p}} \hat{a}_{\vec{p}_1} \hat{c}_{\vec{k}_1 \lambda}^+ \right]$$

(64)

The second term of (42) containing a scalar potential is even quadratic in $A^0$:

$$\hat{I}_7 = \int d^4x_1 \int d^4x_2 \, T\!\left(\phi^+(x_1)A^0(x_1)\phi(x_1)\phi^+(x_2)A^0(x_2)\phi(x_2)\right). \tag{65 a}$$

(65 a) contains a factor of two integrands of the kind of (62 a), but taken at two different times. Thus the time ordering operator must be taken into account. With the same substitutions as in (62) and (64), we obtain the following result:

$$\hat{I}_7 = i\int d^3p_1' \, d^3k_1' \, d^3k_1 \, d^3p_1 \, d^3p_2' \, d^3k_2' \, d^3k_2 \, d^3p_2 \, \frac{1}{2^3(2\pi)^5} \hat{a}^+_{\vec{p}_1'}\hat{a}^+_{\vec{k}_1'} \hat{a}_{\vec{k}_1} \hat{a}_{\vec{p}_1} \hat{a}^+_{\vec{p}_2'}\hat{a}^+_{\vec{k}_2'} \hat{a}_{\vec{k}_2} \hat{a}_{\vec{p}_2} \cdot$$
$$\delta(\omega_{p_2} + \omega_{k_2} - \omega_{p_2'} - \omega_{k_2'} + \omega_{p_1} + \omega_{k_1} - \omega_{p_1'} - \omega_{k_1'})\delta^3(\vec{p}_1' - \vec{p}_1 + \vec{k}_1' - \vec{k}_1)\delta^3(\vec{p}_2' - \vec{p}_2 + \vec{k}_2' - \vec{k}_2) \cdot$$
$$\frac{1}{(\vec{k}_1' - \vec{k}_1)^2} \frac{1}{(\vec{k}_2' - \vec{k}_2)^2} \frac{1}{\omega_{p_2} + \omega_{k_2} - \omega_{p_2'} - \omega_{k_2'} + i\varepsilon} \tag{65 b}$$

The results (52), (54), (55), (57), (62), (64) and (65) substituted into (38) to (42) now enable us to evaluate scattering matrix elements for scattering processes to (and including) the order $e^2$. As two examples, we turn first to the scalar analogue of Compton scattering in order to address then to the scattering of two identical scalar bosons. These two scattering processes can be compared easily with the corresponding results of the well-known scalar QED dealing with the Klein-Gordon equation (3).

**Compton scattering**

For Compton scattering, we need one scalar boson and one photon each in the input and output channel. This means, we have to evaluate the element

$$\langle \hat{S} \rangle := \langle 0 | \hat{c}_{\vec{h}'\mu'} \hat{a}_{\vec{q}'} \hat{S} \, \hat{a}^+_{\vec{q}} \hat{c}^+_{\vec{h}\mu} | 0 \rangle, \tag{66}$$

with the $\hat{S}$-operator (33). Firstly, we realise that the terms based on $\hat{I}_1$ (see (52)) as well as $\hat{I}_6$ (see (64)) must vanish, because the subsequent two elements in the photon operators become zero:

$$\langle 0 | \hat{c}_{\vec{h}'\mu'} \hat{c}_{\vec{k}\lambda} \hat{c}^+_{\vec{h}\mu} | 0 \rangle = \delta_{\vec{k}\vec{h}} \, \delta_{\lambda\mu} \langle 0 | \hat{c}_{\vec{h}'\mu'} | 0 \rangle = 0, \tag{67}$$

$$\langle 0 | \hat{c}_{\vec{h}'\mu'} \hat{c}^+_{\vec{k}\lambda} \hat{c}^+_{\vec{h}\mu} | 0 \rangle = 0.$$

There, we have used the commutation relations (47), the properties of creation and annihilation operators corresponding to those of (22) and (23) as well as abbreviated the delta functional by

$$\delta_{\vec{k}\vec{h}} = \delta^3(\vec{k} - \vec{h}). \tag{68}$$

The terms with $\hat{I}_2$ and $\hat{I}_3$ need

$$\langle 0 | \hat{a}_{\vec{q}'} \hat{a}^+_{\vec{p}_1} \hat{a}_{\vec{p}_2} \hat{a}^+_{\vec{q}} | 0 \rangle = \delta_{\vec{q}'\vec{p}_1} \, \delta_{\vec{p}_2\vec{q}}, \tag{69}$$

whereas a term

$$\langle 0|\hat{a}_{\vec{q}'}\hat{a}^+_{\vec{p}'_2}\hat{a}_{\vec{p}'_1}\hat{a}^+_{\vec{p}_2}\hat{a}_{\vec{p}_1}\hat{a}^+_{\vec{q}}|0\rangle = \delta_{\vec{q}'\vec{p}'_2}\,\delta_{\vec{p}'_1\vec{p}_2}\,\delta_{\vec{p}_1\vec{q}}$$

belongs to $\hat{I}_4$. Here we have used again (22) to (24).

For $\hat{I}_2$, $\hat{I}_3$ and $\hat{I}_4$ the following equations are necessary, too:

$$\langle 0|\hat{c}_{\vec{h}'\mu'}\hat{c}_{\vec{k}'\lambda'}\hat{c}_{\vec{k}\lambda}\hat{c}^+_{\vec{h}\mu}|0\rangle = 0,$$

$$\langle 0|\hat{c}_{\vec{h}'\mu'}\hat{c}^+_{\vec{k}'\lambda'}\hat{c}^+_{\vec{k}\lambda}\hat{c}^+_{\vec{h}\mu}|0\rangle = 0, \qquad (70)$$

$$\langle 0|\hat{c}_{\vec{h}'\mu'}\hat{c}_{\vec{k}'\lambda'}\hat{c}^+_{\vec{k}\lambda}\hat{c}^+_{\vec{h}\mu}|0\rangle = \delta_{\vec{h}'\vec{k}}\,\delta_{\vec{h}\vec{k}'}\,\delta_{\mu'\lambda}\,\delta_{\lambda'\mu},$$

$$\langle 0|\hat{c}_{\vec{h}'\mu'}\hat{c}^+_{\vec{k}'\lambda'}\hat{c}_{\vec{k}\lambda}\hat{c}^+_{\vec{h}\mu}|0\rangle = \delta_{\vec{h}'\vec{k}'}\,\delta_{\vec{h}\vec{k}}\,\delta_{\mu'\lambda'}\,\delta_{\lambda\mu},$$

where we have commutated the creation operators successively to the left and the annihilation operators to the right.

Furthermore, we see that the term based on $\hat{I}_5$ from (62) becomes zero, due to

$$\langle 0|\hat{a}_{\vec{q}'}\hat{a}^+_{\vec{p}'}\hat{a}^+_{\vec{k}'}\hat{a}_{\vec{k}}\hat{a}_{\vec{p}}\hat{a}^+_{\vec{q}}|0\rangle = 0. \qquad (71)$$

A similar result holds true for the term based on $\hat{I}_7$ from (65).

$$\langle 0|\hat{a}_{\vec{q}'}\hat{a}^+_{\vec{p}'_1}\hat{a}^+_{\vec{k}'_1}\hat{a}_{\vec{k}_1}\hat{a}_{\vec{p}_1}\hat{a}^+_{\vec{p}'_2}\hat{a}^+_{\vec{k}'_2}\hat{a}_{\vec{k}_2}\hat{a}_{\vec{p}_2}\hat{a}^+_{\vec{q}}|0\rangle = 0. \qquad (72)$$

Now, we can determine (66) by means of (33):

$$\langle \hat{S}-1 \rangle \approx (-ie)\left(-\tfrac{1}{2}\langle\hat{I}_1\rangle + \langle\hat{I}_5\rangle\right) + \tfrac{1}{2}(-ie)^2\left(i\langle\hat{I}_2\rangle - i\tfrac{1}{4}\langle\hat{I}_3\rangle + \tfrac{1}{4}\langle\hat{I}_4\rangle - \tfrac{1}{2}\langle\hat{I}_6\rangle + \langle\hat{I}_7\rangle\right). \qquad (73)$$

For Compton scattering, $\langle\hat{I}_1\rangle$, $\langle\hat{I}_2\rangle$, $\langle\hat{I}_3\rangle$, $\langle\hat{I}_4\rangle$, $\langle\hat{I}_5\rangle$, $\langle\hat{I}_6\rangle$ and $\langle\hat{I}_7\rangle$ can be evaluated explicitly:

$$\langle\hat{I}_1\rangle = \langle\hat{I}_5\rangle = \langle\hat{I}_6\rangle = \langle\hat{I}_7\rangle = 0, \qquad (74)$$

$$\langle\hat{I}_2\rangle = \frac{\delta^4(q+h-q'-h')}{\sqrt{2\pi}\sqrt{\tilde{\omega}_{h'}\tilde{\omega}_h}\,\omega_q}\,\vec{\varepsilon}(\vec{h},\mu)\cdot\vec{\varepsilon}(\vec{h}',\mu'), \qquad (75)$$

$$\langle\hat{I}_3\rangle = \frac{\delta^4(q+h-q'-h')}{(2\pi)^2\,2\sqrt{\tilde{\omega}_{h'}\tilde{\omega}_h}\,\omega_q}\left[\frac{1}{\omega^2_{\vec{q}+\vec{h}}}\vec{\varepsilon}(\vec{h}',\mu')\cdot(2\vec{q}+\vec{h}-\vec{h}')\vec{\varepsilon}(\vec{h},\mu)\cdot(2\vec{q}+\vec{h}) + \right.$$

$$\left. \frac{1}{\omega^2_{\vec{q}-\vec{h}'}}\vec{\varepsilon}(\vec{h},\mu)\cdot(2\vec{q}-\vec{h}'+\vec{h})\vec{\varepsilon}(\vec{h}',\mu')\cdot(2\vec{q}-\vec{h}')\right] \qquad (76)$$

$$\langle \hat{I}_4 \rangle = \frac{\delta^4(q+h-q'-h')}{(2\pi)^2 \sqrt{\tilde{\omega}_{h'}\tilde{\omega}_h}\,\omega_q} \left[ \frac{1}{\omega_{\vec{q}-\vec{h}'}} \frac{\vec{\varepsilon}(\vec{h},\mu)\cdot(2\vec{q}'-\vec{h})\,\vec{\varepsilon}(\vec{h}',\mu')\cdot(2\vec{q}-\vec{h}')}{\omega_{\vec{q}} - \tilde{\omega}_{\vec{h}'} - \omega_{\vec{q}-\vec{h}'} + i\varepsilon} + \right.$$
$$\left. \frac{1}{\omega_{\vec{q}+\vec{h}}} \frac{\vec{\varepsilon}(\vec{h}',\mu')\cdot(2\vec{q}'+\vec{h})\,\vec{\varepsilon}(\vec{h},\mu)\cdot(2\vec{q}+\vec{h})}{\omega_{\vec{q}} + \tilde{\omega}_{\vec{h}} - \omega_{\vec{q}+\vec{h}} + i\varepsilon} \right]$$
(77)

The two terms $\langle \hat{I}_2 \rangle$ and $\langle \hat{I}_4 \rangle$ in (73) resemble the three terms in the corresponding formula of Compton scattering for scalar bosons, but this time being based on the Klein-Gordon equation (3) (see e.g. [6,8]):

$$\langle \hat{S} - 1 \rangle = \frac{(-ie)^2}{(2\pi)^2 \sqrt{2\tilde{\omega}_{h'} 2\tilde{\omega}_h 2\omega_{q'} 2\omega_q}} \delta^4(q+h-q'-h') \cdot$$
$$\left[ \varepsilon(h,\mu)\cdot(2q+h)\frac{i}{(q+h)^2 - m^2} \varepsilon(h',\mu')\cdot(2q'+h') \right.$$
$$+ \varepsilon(h,\mu)\cdot(2q'-h)\frac{i}{(q-h')^2 - m^2} \varepsilon(h',\mu')\cdot(2q-h')$$
$$\left. - 2i\,\varepsilon(h,\mu)\cdot\varepsilon(h',\mu') \right]$$
(78)

where $\varepsilon$ is the 4-dimensional generalisation of the three dimensional polarisation vector $\vec{\varepsilon}$ used so far. In (78), the first two terms correspond to $\langle \hat{I}_4 \rangle$. We realise that these terms in (78) look very much like those in (77), but also that especially the propagators of both theories are a bit different. (75) resembles the third term in (78). On the other hand, $\langle \hat{I}_3 \rangle$ in (76) could also be regarded as the analogue of the first two terms in (78) – at least after having used the Coulomb gauge condition (49) in (76).
The first two terms in (78) vanish, if we choose the incoming scalar boson to be at rest,

$$q = (q^0, \vec{q}) := (m, \vec{0}),$$
(79)

and want to have transversally polarised photons in this laboratory system,

$$\varepsilon(h,\mu)\cdot q = \varepsilon(h',\mu')\cdot q = 0,$$
(80)

and use the Lorentz gauge condition

$$\varepsilon(h,\mu)\cdot h = \varepsilon(h',\mu')\cdot h' = 0.$$
(81)

Accordingly, $\langle \hat{I}_3 \rangle$ and $\langle \hat{I}_4 \rangle$ in (73) vanish too, if we adopt (79) again and use the analogue of (80),

$$\vec{\varepsilon}(\vec{h},\mu)\cdot \vec{q} = \vec{\varepsilon}(\vec{h}',\mu')\cdot \vec{q} = 0,$$
(82)

as well as apply the Coulomb gauge condition (49).
That is, under these conditions in both versions of scalar QED, only one term remains (i.e. (75) in (73) and, accordingly, the third term in (78)) which is a relativistically generalised version of the matrix element from which the well-known Thomson scattering cross section can be evaluated.
Even though the propagators in both theories are rather different, for this example of scattering process, the results do not seem to differ very much from each other in the laboratory system chosen above. Therefore the question arises, whether this is also the case for further scattering processes. To this end, we are going to investigate what happens, if two identical scalar bosons interact with each other.

**Scattering of two identical scalar bosons**

If we want to calculate matrix elements of the *S*-operator (33) for the scattering of two identical scalar bosons, we need two scalar bosons in the input channel and two in the output channel:

$$\left\langle \hat{S} \right\rangle := \left\langle 0 \left| \hat{a}_{\vec{q}_2'} \hat{a}_{\vec{q}_1'} \hat{S} \, \hat{a}_{\vec{q}_2}^+ \hat{a}_{\vec{q}_1}^+ \right| 0 \right\rangle. \tag{83}$$

We can reuse (73), but have to evaluate $\left\langle \hat{I}_1 \right\rangle$, $\left\langle \hat{I}_2 \right\rangle$, $\left\langle \hat{I}_3 \right\rangle$, $\left\langle \hat{I}_4 \right\rangle$, $\left\langle \hat{I}_5 \right\rangle$, $\left\langle \hat{I}_6 \right\rangle$ and $\left\langle \hat{I}_7 \right\rangle$ again. Firstly, we recognise that due to (52) and the analogue of (22) and (23) for the photon operators

$$\left\langle \hat{I}_1 \right\rangle = 0, \tag{84}$$

is valid. Furthermore, we need

$$\left\langle 0 \left| \hat{c}_{\vec{k}_1 \lambda_1} \hat{c}_{\vec{k}_2 \lambda_2}^+ \right| 0 \right\rangle = \delta_{\vec{k}_1 \vec{k}_2} \, \delta_{\lambda_1 \lambda_2} \tag{85}$$

for calculating $\left\langle \hat{I}_2 \right\rangle$ with the help of (54), whereas all the other photon operator terms therein vanish. The same holds true for $\left\langle \hat{I}_3 \right\rangle$ with (55) and $\left\langle \hat{I}_4 \right\rangle$ with (57).

As far as the scalar boson operators are concerned, for the evaluation of $\left\langle \hat{I}_2 \right\rangle$ and $\left\langle \hat{I}_3 \right\rangle$

$$\left\langle 0 \left| \hat{a}_{\vec{q}_2'} \hat{a}_{\vec{q}_1'} \hat{a}_{\vec{p}_2}^+ \hat{a}_{\vec{p}_1} \hat{a}_{\vec{q}_2}^+ \hat{a}_{\vec{q}_1}^+ \right| 0 \right\rangle = 0 .$$

Hence, these two terms,

$$\left\langle \hat{I}_2 \right\rangle = 0, \tag{86}$$
$$\left\langle \hat{I}_3 \right\rangle = 0,$$

are zero, too. For the evaluation of the non-vanishing term $\left\langle \hat{I}_4 \right\rangle$, the following result is useful:

$$\left\langle 0 \left| \hat{a}_{\vec{q}_2'} \hat{a}_{\vec{q}_1'} \hat{a}_{\vec{p}_2'}^+ \hat{a}_{\vec{p}_1'} \hat{a}_{\vec{p}_2}^+ \hat{a}_{\vec{p}_1} \hat{a}_{\vec{q}_2}^+ \hat{a}_{\vec{q}_1}^+ \right| 0 \right\rangle = \delta_{\vec{q}_1 \vec{p}_1} \delta_{\vec{q}_2 \vec{p}_1'} \delta_{\vec{q}_2' \vec{p}_2'} \delta_{\vec{q}_1' \vec{p}_2} + \delta_{\vec{q}_1 \vec{p}_1} \delta_{\vec{q}_2 \vec{p}_1'} \delta_{\vec{q}_2' \vec{p}_2} \delta_{\vec{q}_1' \vec{p}_2'} + \\ \delta_{\vec{q}_1 \vec{p}_1'} \delta_{\vec{q}_2 \vec{p}_1} \delta_{\vec{q}_2' \vec{p}_2'} \delta_{\vec{q}_1' \vec{p}_2} + \delta_{\vec{q}_1 \vec{p}_1'} \delta_{\vec{q}_2 \vec{p}_1} \delta_{\vec{q}_2' \vec{p}_2} \delta_{\vec{q}_1' \vec{p}_2'} . \tag{87}$$

With (85), (87) and the invariance of (48) under the transformation $\vec{k} \to -\vec{k}$, we get

$$\langle \hat{S} - 1 \rangle = \frac{-ie^2 \delta^4(q_1 + q_2 - q_1' - q_2')}{8(2\pi)^2 \omega_{q_1} \omega_{q_2}} \left[ \sum_\lambda \frac{(\vec{q}_2 + \vec{q}_2') \cdot \vec{\varepsilon}(\vec{q}_1 - \vec{q}_1', \lambda) \vec{\varepsilon}(\vec{q}_1 - \vec{q}_1', \lambda) \cdot (\vec{q}_1 + \vec{q}_1')}{\tilde{\omega}_{\vec{q}_1 - \vec{q}_1'}} \right.$$

$$\cdot \left( \frac{1}{\omega_{q_1} - \omega_{q_1'} - \tilde{\omega}_{\vec{q}_1 - \vec{q}_1'} + i\varepsilon} - \frac{1}{\omega_{q_1} - \omega_{q_1'} + \tilde{\omega}_{\vec{q}_1 - \vec{q}_1'} - i\varepsilon} \right) +$$

$$\sum_\lambda \frac{(\vec{q}_2 + \vec{q}_1') \cdot \vec{\varepsilon}(\vec{q}_1 - \vec{q}_2', \lambda) \vec{\varepsilon}(\vec{q}_1 - \vec{q}_2', \lambda) \cdot (\vec{q}_1 + \vec{q}_2')}{\tilde{\omega}_{\vec{q}_1 - \vec{q}_2'}}$$

$$\left. \cdot \left( \frac{1}{\omega_{q_1} - \omega_{q_2'} - \tilde{\omega}_{\vec{q}_1 - \vec{q}_2'} + i\varepsilon} - \frac{1}{\omega_{q_1} - \omega_{q_2'} + \tilde{\omega}_{\vec{q}_1 - \vec{q}_2'} - i\varepsilon} \right) \right] + \text{terms in } A^0 \quad (88\ a)$$

Here, we have omitted the terms $\langle \hat{I}_5 \rangle$, $\langle \hat{I}_6 \rangle$ and $\langle \hat{I}_7 \rangle$ containing terms in the scalar potential $A^0$ which are regarded later.

(88 a) can be compared directly with the corresponding result of the Klein-Gordon equation (3) (see e.g. [9]):

$$\langle \hat{S} - 1 \rangle = \frac{ie^2 \delta^4(q_1 + q_2 - q_1' - q_2')}{4(2\pi)^2 \sqrt{\omega_{q_1} \omega_{q_2} \omega_{q_1'} \omega_{q_2'}}} \left[ \frac{(q_1 + q_1') \cdot (q_2 + q_2')}{(q_1 - q_1')^2} + \frac{(q_1 + q_2') \cdot (q_2 + q_1')}{(q_1 - q_2')^2} \right], \quad (89)$$

if we use (48) and leave out the term $\varepsilon$ approaching zero in (88):

$$\langle \hat{S} - 1 \rangle = \frac{-ie^2 \delta^4(q_1 + q_2 - q_1' - q_2')}{4(2\pi)^2 \omega_{q_1} \omega_{q_2}} \left[ \frac{(\vec{q}_1 + \vec{q}_1') \cdot (\vec{q}_2 + \vec{q}_2')}{(q_1 - q_1')^2} + \frac{(\vec{q}_1 + \vec{q}_2') \cdot (\vec{q}_2 + \vec{q}_1')}{(q_1 - q_2')^2} \right.$$

$$\left. - \frac{(\vec{q}_2 + \vec{q}_2') \cdot (\vec{q}_1 - \vec{q}_1')(\vec{q}_1 - \vec{q}_1') \cdot (\vec{q}_1 + \vec{q}_1')}{(q_1 - q_1')^2 (\vec{q}_1 - \vec{q}_1')^2} - \frac{(\vec{q}_2 + \vec{q}_1') \cdot (\vec{q}_1 - \vec{q}_2')(\vec{q}_1 - \vec{q}_2') \cdot (\vec{q}_1 + \vec{q}_2')}{(q_1 - q_2')^2 (\vec{q}_1 - \vec{q}_2')^2} \right]$$

$+ \text{terms in } A^0$

(88 b)

The first two terms in (88 b) correspond to (89) which contains the photon propagator in the Feynman gauge. But since we use a Coulomb gauge, only the space-like components of the linear 4-momenta appear in the numerators of the first two terms in (88 b). Moreover, in the second line of (88 b) two additional terms are present which can be reformulated by means of the delta distribution in (88 b):

$$- \frac{(\vec{q}_2'^2 - \vec{q}_2^2)(\vec{q}_1^2 - \vec{q}_1'^2)}{(q_1 - q_1')^2 (\vec{q}_1 - \vec{q}_1')^2} - \frac{(\vec{q}_1'^2 - \vec{q}_2^2)(\vec{q}_1^2 - \vec{q}_2'^2)}{(q_1 - q_2')^2 (\vec{q}_1 - \vec{q}_2')^2}. \quad (90)$$

The structure of (88 b) is the same as that of (89). We have two terms: in the second term, the momenta of the two scalar bosons in the output channel have been exchanged compared to the first term.

Now we address to the terms in the scalar potential $A^0$ in (88). The Coulomb term $\langle \hat{I}_5 \rangle$ contains a term

$$\langle 0 | \hat{a}_{\vec{q}_2'} \hat{a}_{\vec{q}_1'} \hat{a}_{\vec{p}'}^+ \hat{a}_{\vec{k}'}^+ \hat{a}_{\vec{k}} \hat{a}_{\vec{p}} \hat{a}_{\vec{q}_2}^+ \hat{a}_{\vec{q}_1}^+ | 0 \rangle = \delta_{\vec{p}\vec{q}_2} \delta_{\vec{k}\vec{q}_1} \delta_{\vec{q}_1'\vec{p}'} \delta_{\vec{q}_2'\vec{k}'} - \delta_{\vec{p}\vec{q}_2} \delta_{\vec{k}\vec{q}_1} \delta_{\vec{q}_1'\vec{k}'} \delta_{\vec{q}_2'\vec{p}'}$$

$$- \delta_{\vec{k}\vec{q}_2} \delta_{\vec{p}\vec{q}_1} \delta_{\vec{q}_1'\vec{p}'} \delta_{\vec{q}_2'\vec{k}'} - \delta_{\vec{k}\vec{q}_2} \delta_{\vec{p}\vec{q}_1} \delta_{\vec{q}_1'\vec{k}'} \delta_{\vec{q}_2'\vec{p}'}.$$

Therefore, $\langle \hat{I}_5 \rangle$ yields

$$\hat{I}_5 = \frac{1}{(2\pi)^2} \delta^4(q_1' + q_2' - q_1 - q_2) \left[ \frac{1}{(\vec{q}_1' - \vec{q}_1)^2} - \frac{1}{(\vec{q}_2' - \vec{q}_1)^2} \right]. \tag{91}$$

The term $\langle \hat{I}_6 \rangle$ vanishes, since the annihilation operators for photons in (64) act directly on the vacuum states.

On the other hand, $\langle \hat{I}_7 \rangle$ becomes a rather lengthy because of

$$\langle 0 | \hat{a}_{\vec{q}_2'} \hat{a}_{\vec{q}_1'} \hat{a}^+_{\vec{p}_1'} \hat{a}^+_{\vec{k}_1'} \hat{a}_{\vec{k}_1} \hat{a}_{\vec{p}_1} \hat{a}^+_{\vec{p}_2'} \hat{a}^+_{\vec{k}_2'} \hat{a}_{\vec{k}_2} \hat{a}_{\vec{p}_2} \hat{a}^+_{\vec{q}_2} \hat{a}^+_{\vec{q}_1} | 0 \rangle$$

$$= \left( \delta_{\vec{p}_2 \vec{q}_2} \delta_{\vec{p}_1 \vec{p}_2'} \delta_{\vec{k}_2 \vec{q}_1} \delta_{\vec{k}_1 \vec{k}_2'} \delta_{\vec{q}_1' \vec{p}_1'} \delta_{\vec{q}_2' \vec{k}_1'} - (\vec{p}_1' \leftrightarrow \vec{k}_1') \right)$$

$$- \left( \delta_{\vec{p}_2 \vec{q}_2} \delta_{\vec{k}_2 \vec{q}_1} \delta_{\vec{k}_1 \vec{p}_2'} \delta_{\vec{p}_1 \vec{k}_2'} \delta_{\vec{q}_1' \vec{p}_1'} \delta_{\vec{q}_2' \vec{k}_1'} - (\vec{p}_1' \leftrightarrow \vec{k}_1') \right)$$

$$- \left( \delta_{\vec{p}_2 \vec{q}_2} \delta_{\vec{k}_2 \vec{q}_2} \delta_{\vec{p}_2 \vec{q}_1} \delta_{\vec{p}_1 \vec{p}_2'} \delta_{\vec{q}_1' \vec{p}_1'} \delta_{\vec{q}_2' \vec{k}_1'} \delta_{\vec{k}_1 \vec{k}_2'} - (\vec{p}_1' \leftrightarrow \vec{k}_1') \right)$$

$$+ \left( \delta_{\vec{p}_2 \vec{q}_2} \delta_{\vec{k}_2 \vec{q}_2} \delta_{\vec{p}_2 \vec{q}_1} \delta_{\vec{k}_1 \vec{p}_2'} \delta_{\vec{p}_1 \vec{k}_2'} \delta_{\vec{q}_1' \vec{p}_1'} \delta_{\vec{q}_2' \vec{k}_1'} - (\vec{p}_1' \leftrightarrow \vec{k}_1') \right)$$

where $(\vec{p}_1' \leftrightarrow \vec{k}_1')$ denotes the same term as the immediately preceding one, but with exchanged momenta $\vec{p}'$ and $\vec{k}_1'$, respectively. Thus, $\langle \hat{I}_7 \rangle$ gives

$$\langle \hat{I}_7 \rangle = \frac{i}{8(2\pi)^2} \delta^4(q_1' + q_2' - q_1 - q_2) \int \frac{d^3 p}{(2\pi)^3} \frac{1}{\omega_{\vec{q}_1} + \omega_{\vec{q}_2} - \omega_{\vec{p}} - \omega_{\vec{q}_1 + \vec{q}_2 - \vec{p}}} \cdot$$

$$\left[ \frac{1}{\tilde{\omega}^2_{\vec{p} - \vec{q}_1'} \tilde{\omega}^2_{\vec{p} + \vec{q}_2}} - \frac{1}{\tilde{\omega}^2_{\vec{p} - \vec{q}_2'} \tilde{\omega}^2_{\vec{p} + \vec{q}_2}} - \frac{1}{\tilde{\omega}^2_{\vec{p} - \vec{q}_1'} \tilde{\omega}^2_{\vec{p} - \vec{q}_1}} + \frac{1}{\tilde{\omega}^2_{\vec{p} - \vec{q}_2'} \tilde{\omega}^2_{\vec{p} - \vec{q}_1}} \right] \tag{92}$$

Hence in place of the time-like components (i.e. the energy terms) of the linear 4-momenta in the numerators of (89) derived from the Klein-Gordon equation, several terms arise: (90), (91) and (92). But (91) and (the non relativistic limit of) (92) would also have appeared, if we had started from the non-relativistic Schrödinger equation. Thus, these terms state the fact that the equation used for obtaining (88) (together with (90), (91) and (92)) is Schrödinger equation like.

**Conclusions and outlook**

For scalar bosons, we could see that it is possible to describe scattering processes by means of a square root operator equation being coupled to an electromagnetic field. We achieved this by splitting off a factor in the shape of the free square root operator from the equation and by a series expansion of a remaining square root factor containing terms in the electromagnetic vector potential and powers of the (inverse) free square root operator. The latter ones could be given an integral representation.
Having quantised the fields involved, we could evaluate the scattering matrix elements for Compton scattering and for the scattering of two identical bosons (to – and including – the quadratic order of $e$) which, on the one hand, resembled the results derived with the help of the corresponding Klein-Gordon equation and, on the other hand, the results one would have obtained with a non-relativistic Schrödinger equation.
Of course, now several questions arise, e.g.:

- Can we formulate Feynman rules at all for our non-local scalar QED?

- Do divergent terms appear and, if yes, can a renormalisation procedure be found?
- Can the results of this non-local QED be confirmed (or refuted) by experiments?

To the first question: if one tries to formulate Feynman rules, one must be aware that in each step of the approximation procedure, we have to expand the first square root factor (in the second term of) (7) to the desired order in $e$. Thus, the Hamiltonian we are using within that procedure must be adapted in each order of $e$ of the approximation. Therefore we conclude that even if it were possible to formulate Feynman rules, they would be much more complicated than those for the Klein-Gordon theory. This is the price we have to pay for non-locality. But without Feynman rules, it is not very easy to analyse the renormalisability of that non-local scalar QED either.

The answer to the last question listed above is negative due to a lack of elementary spinless bosons in nature. But that question would be sensible, if we had a corresponding non-local theory for spin-1/2 particles. We could even make a guess, how this theory would look like: for free spin-1/2 particles, the wave functions in (2) should be 2-spinors. It would also be possible to couple this equation to an electromagnetic field, but the usual minimal coupling scheme does not work. Instead we would have to postulate an equation:

$$i\partial_t \phi = \hat{H}'_\mp \phi, \tag{93}$$

with the Hamilton operator

$$\hat{H}'_\mp = \sqrt{m^2 + (\hat{\vec{p}} - e\vec{A})^2 - e\vec{\sigma} \cdot (\vec{B} \mp i\vec{E})} + eA^0, \tag{94}$$

which contains an additional term with the Pauli matrices $\vec{\sigma}$ and the magnetic field $\vec{B}$ as well as the electric field $\vec{E}$ under the square root. For this equation, it has already been shown that it can reproduce the gyromagnetic factor of 2 for the electron as well as, when being applied to a hydrogen atom, that it can reproduce correct binding energies of the electron at least to (and including) the quadratic order in $e^2$ (see [12]). We can apply the same approximation procedure to this equation as we have presented here for the scalar case. But of course, additional difficulties emerge: the Hamiltonian corresponding to the one shown in (25) should be Hermitian. And at least for the free spin-1/2 case, that Hamiltonian should be relativistically invariant. This means, that for the free case, (25) should be a Lorentz scalar with respect to the spin. To this end, we have to replace $\phi$ and $\phi^+$ therein by combinations of a mixture of left and right handed 2-spinors $\phi_L$ and $\phi_R$, respectively, and their Hermitian conjugates, because $\phi_L^+ \phi_R$ and $\phi_R^+ \phi_L$ are Lorentz scalars. Therefore, (25) could be replaced either by

$$\hat{H} = \tfrac{1}{2}\phi_R^+ \hat{H}'_\mp \phi_L + \tfrac{1}{2}\phi_L^+ \hat{H}'_\pm \phi_R \tag{95}$$

or by

$$\hat{H} = \tfrac{1}{2}\phi_R^+ \hat{H}'_\pm \phi_L + \tfrac{1}{2}\phi_L^+ \hat{H}'_\mp \phi_R \tag{96}$$

so that the Hamiltonians (95) and (96) become Hermitian, because of the property

$$\hat{H}'_\mp = \hat{H}'^{+}_\pm. \tag{97}$$

The results of a QED based on (95) or (96) could then be compared with the ones based on a corresponding Dirac equation.

The author does not know, whether such an approach has already been performed for the spin-1/2 case or for the here presented spinless case. He does not know either, if an application for it can be found, where e.g. non-local properties are indispensable. But it seems to him that from the technical point of view, square root operator equations coupled to an electromagnetic field like (4) (or maybe even like (93) together with (94)) can be handled within the framework of a quantum field theory. Such equations were given up quite early in the history of quantum mechanics for several good reasons, e.g. due to their lack of relativistic invariance (see [13]), their non-local character accompanied by the difficulty of finding an appropriate mathematical interpretation and description, respectively (see e.g. [4,5,6]). While the first reason mentioned remains still valid, from today's perspective, those non-local properties might not be refused as vehemently as in the past (e.g. when one looks out for approximations of the so called Bethe-Salpeter equation [11]). The author hopes to have shown, that at least answers to the question of possible descriptions of such non-local square root operator equations can be found.